\documentclass[10pt,sigconf,letterpaper]{acmart}

\usepackage{amsmath}
\usepackage{algorithmic}
\usepackage{graphicx}
\usepackage{textcomp}
\usepackage{xcolor}
\usepackage{listings}
\usepackage{makecell}
\usepackage{pdfpages}
\usepackage{xurl}
\usepackage{todonotes}
\usepackage{seqsplit}
\usepackage{comment}
\usepackage{booktabs}
\usepackage{makecell} 
\usepackage{fancyvrb}

\AtBeginDocument{%
  \providecommand\BibTeX{{%
    \normalfont B\kern-0.5em{\scshape i\kern-0.25em b}\kern-0.8em\TeX}}}

\copyrightyear{2021}

\acmSubmissionID{418}

\setcopyright{none}
\renewcommand\footnotetextcopyrightpermission[1]{}

\begin{document}

\title{Large Scale Measurement on the Adoption of Encrypted DNS}


\author{Sebastián García}
\email{garciseb@fel.cvut.cz}
\orcid{0000-0001-6238-9910}
\affiliation{%
  \institution{Faculty of Electrical Engineering \\ Czech Technical University in Prague}
  \streetaddress{Technická 2}
  \city{Prague}
  \country{Czech Republic}
  \postcode{160 00}
}

\author{Karel Hynek}
\email{hynekkar@fit.cvut.cz}
\orcid{0000-0002-8281-618X}
\affiliation{%
  \institution{Faculty of Information Technology \\ Czech Technical University in Prague \\ and \\ CESNET z.s.p.o.}
  \streetaddress{Thákurova 9}
  \city{Prague}
  \country{Czech Republic}
  \postcode{160 00}
}

\author{Dmtrii Vekshin}
\email{dmitrii.vekshin@avast.com}
\affiliation{%
  \institution{Avast s.r.o.}
  \streetaddress{Pikrtova 1737}
  \city{Prague}
  \country{Czech Republic}
  \postcode{140 00}
}

\author{Tomáš Čejka}
\email{cejkat@cesnet.cz}
\orcid{0000-0001-7794-9511}
\affiliation{%
  \institution{CESNET z.s.p.o.}
  \streetaddress{Zikova 4}
  \city{Prague}
  \country{Czech Republic}
  \postcode{160 00}
}

\author{Armin Wasicek}
\email{armin.wasicek@avast.com}
\affiliation{%
  \institution{Avast s.r.o.}
  \streetaddress{Pikrtova 1737}
  \city{Prague}
  \country{Czech Republic}
  \postcode{140 00}
}

\begin{abstract}
Several encryption proposals for DNS have been presented since 2016, but their adoption was not comprehensively studied yet. This research measured the current adoption of DoH (DNS over HTTPS), DoT (DNS over TLS), and DoQ (DNS over QUIC) for five months at the beginning of 2021 by three different organizations with global coverage. By comparing the total values, amount of requests per user, and the seasonality of the traffic, it was possible to obtain the current adoption trends. Moreover, we actively scanned the Internet for still-unknown working DoH servers and we compared them with a novel curated list of well-known DoH servers. We conclude that despite growing in 2020, during the first five months of 2021 there was statistically significant evidence that the average amount of Internet traffic for DoH, DoT and DoQ remained stationary. However, we found that the amount of, still unknown and ready to use, DoH servers grew 4 times. These measurements suggest that even though the amount of encrypted DNS is currently not growing, there may probably be more connections soon to those unknown DoH servers for benign and malicious purposes.
\end{abstract}

\keywords{DoH, DoT, Encrypted DNS, Network Measurement, Network Trends}

\begin{teaserfigure}
  \centering
  \includegraphics[width=\textwidth]{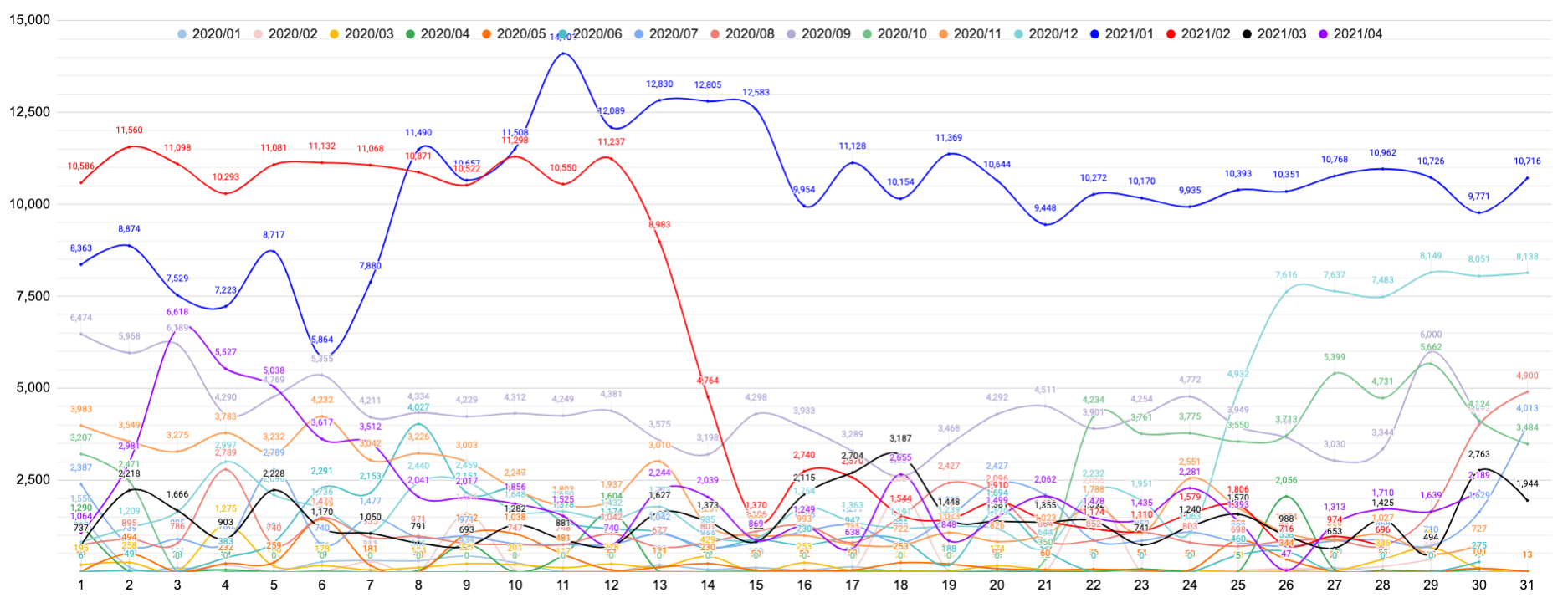}
  \caption{Sum of DoH, DoT, DoQ flows for Organization 2 for 16 months, x-axis are the days of the month.}
  \Description{Comp}
  \label{fig:teaser}
\end{teaserfigure}

\maketitle
\pagestyle{plain}


\section{Introduction}

The Domain Name System (DNS) is a core service that plays an essential role in network communication, but it does not meet the current requirements of privacy and security. In the last years there has been an increasing amount of proposed protocols to enhance its privacy and confidentiality via encryption~\cite{dnssecurity}. This trend started around 2014~\cite{Hu2014}, with protocols like DNS over HTTPS (DoH), DNS over TLS (DoT), or DNS over QUIC (DoQ).




There has been previous measurements of the prevalence of encrypted DNS in the Internet focused on specific protocols of encrypted DNS ~\cite{Doan2021,lu2019}. However, there is a lack of real-world observations of the volume of encrypted DNS traffic, and there is no study of the long term trends on its adoption. Moreover, the amount of DoH servers seems to be small and highly controlled by organizations.


This paper presents the first comprehensive measurement, comparison and analysis of the volume of DoH, Dot and DoQ traffic from the perspective of three large organizations: a) the backbone infrastructure of an Internet Service Provider, b) the network infrastructure of a large University, and c) the traffic of millions of endpoints from a global security company. 

The methodology of our analysis has two main parts. First, a measurement on the total amount of DoH servers on the Internet via a global scan and verification. Second, a deep statistical analysis on the trends, stationarity, patterns and ratios between the three organizations during the first five months of 2021. The analysis is completed by a comparison of the trends with the traffic from 2020. All trends and ratios are statistically tested for similarities and the stationarity of the time series is verified with the Augmented Dickey–Fuller test.

The results show that despite a growing increase of encrypted DNS traffic in 2020, the first five months of 2021 show a statistically significant stationarity of the traffic. With some differences for DoH and DoT traffic, these protocols are used to encrypt up to 0.01\% of the total DNS traffic. Regarding the unknown DoH servers on the Internet, results show that there are at least 4 times more DoH servers working than the most comprehensive and current list of well-known DoH servers. These services could be easily exploited by security threats. 

The contributions of this paper are: 
\begin{itemize}
    \item A comprehensive dataset of well-known DoH providers.
    \item A dataset of unknown and unpublished DoH servers by a global Internet scan.
    \item A measurement and comparison of DoH, DoT and DoQ traffic in three large organizations.
    \item A new Nmap NSE script tool to verify DoH servers.
\end{itemize}

\section{Related Work}
DNS encryption is a relatively novel approach to improve user's privacy. IETF standardized DNS over TLS (DoT) in 2016 as RFC~7858~\cite{rfc7858}, followed by DNS over HTTPS (DoH) in 2018 as RFC~8484~\cite{rfc8484}, and the third approach, DNS over QUIC (DoQ), is still an RFC draft~\cite{huitema-dprive-dnsoquic}. Since DNS encryption modifies the usage of one of the oldest and essential protocols on the Internet,  many research studies focused on the consequences of large-scale DNS encryption. 

Borgolte et al.~\cite{borgolte2019dohreshaping} provided a general discussion about DoH and several areas such as performance, security, and privacy. Similarly, Fidler et al.~\cite{fidler2019} presented challenges that network operators are going to face due to DNS encryption. Both studies concluded that large deployments of privacy-preserving DNS will be a problem for users' security since automated network security tools rely on unencrypted DNS. This phenomenon is further researched by Bumanglag et al.~\cite{bumanglang2020}, which analyzed DoH traffic and even discussed the challenges related to its detection, since DoH uses a well-known port allocated for HTTPS communication (443/TCP). The Bumanglag paper also mentioned that the cybersecurity community suggests allowing HTTPS traffic to pass only through inspection proxies that would decrypt and encrypt all HTTPS traffic.

The detection of DoH in different traffic was studied in multiple papers. There are currently two main approaches in DoH detection. The first approach is the detection of regular behavior patterns by Hjelm et al.~\cite{DoHDEtection-Hjelm}. They claim that by using the Real Intelligence Threat Analytics (RITA) framework, they were able to identify behavioral patterns of DoH and successfully detect them. The second approach employs Machine Learning principles and was introduced by Vekshin et al.~\cite{Vekshin2020}. Their AdaBoost decision tree classifier was able to detect DoH connections with an accuracy of more the 99\,\%. A limitation was that their classifier worked only with DoH created by web browsers and cannot detect single DoH queries.

Malicious DoH studies were limited only to data exfiltration via DNS. MontazeriShatoori et al.~\cite{MontazeriShatoori2020} created a classifier that can distinguish between a DoH tunnel and regular DoH with outstanding precision. Similar results were achieved on the same dataset by Singh et al.~\cite{Singh2020} and Banadaki~\cite{banadaki2020detecting}.

Several studies~\cite{bushart2019padding,doh2019encrypted,hynek2020,houser2019} also challenge the privacy-preserving characteristics of encrypted DNS. Studies from Siby et al.~\cite{doh2019encrypted} and Bushart et al.~\cite{bushart2019padding} brought interesting results by performing very accurate website fingerprinting using only DoH and DoT traffic. Even though their results showed a limitation of encryption, their proposed attack is difficult to reproduce in a real environment but it could be fixed by proper payload padding. Nevertheless, the main reason for privacy concerns is still the transition of visibility from local DNS resolvers controlled by Internet Service Providers (ISPs) to global DoH service providers that centralize the entire process of domain name translation.
 
The previously mentioned studies focused mainly on encrypted DNS characteristics, weaknesses, and detection possibilities. However, none of them dealt with the adoption and actual popularity of DNS encryption. Deccio et al.~\cite{Deccio2019} studied in 2019 the adoption of DoT and DoH by open resolvers. Their result shows that the adoption is very poor. From around 1.2 million open resolvers, only 1,747 (0.15\,\%) supported DoT, and only 9 supported DoH. Doan et al.~\cite{Doan2021} performed a similar measurement focusing on DoT in 2020 and found 2,151 open resolvers supporting DoT, which shows an increasing trend in resolvers adoption. However, these data show only the adoption across the open DNS resolvers and do not consider specialized DoT/DoH proxies. 

Furthermore, Doan et al.~\cite{Doan2021} and Lu et al.~\cite{lu2019} also measured trends in encrypted DNS usage. Both studies used a dataset from ISP backbone lines collected in 2019, when encrypted DNS was not enabled by default in browsers, and it was much less popular. Additionally, both studies focused mainly on measuring the prevalence of DoT, and the DoH measurement was very limited. Last but not least, their studies did not consider DNS over QUIC. 

We are not aware of any previous work that tried to enumerate all DoH resolvers to infer the encrypted DNS prevalence. Additionally, none of the previous studies measured the prevalence of DNS over QUIC protocol.


\section{DNS Encryption Approaches}
The high-level concepts of DNS were introduced in RFC 1034~\cite{rfc1034} at the beginning of the Internet itself. Since then, DNS became one of the essential protocols on the Internet designed to translate human-friendly domains to IP addresses. The original DNS protocol standardized in RFC 1035~\cite{rfc1035} does not consider security and privacy at all; therefore, it can leak a lot of private information or be an essential part of the attack vector (e.g., in the form of rogue DNS). 

Studies like~\cite{HERRMANN201317,Klein2019} showed the possibility of user tracking over DNS, even bypassing the private mode in browsers. Additionally, large-scale DNS surveillance programs such as QUANTUMDNS and MORECOWBELL operated by governmental agencies~\cite{Grothoof2017} triggered concerns over user privacy in the broad public. The natural response to these concerns is the privacy-preserving encrypted DNS protocol. In this section, we provide a brief description of the four most important DNS encryption approaches, and their summary is shown in Table~\ref{tab:EncryptedDNS-summary}.


\begin{table}[t]
  \caption{Summary of the most important Encrypted DNS method}
  \label{tab:EncryptedDNS-summary}
  \begin{tabular}{l c c c}
    \toprule
    & DoT & DoH  & DoQ \\
    \midrule
    L4 Protocol & TCP/UDP      & TCP           & UDP  \\
    Port        & 853          & 443           & 784  \\
    Protocol    & TLS/DTSL     & HTTPS         & QUIC \\
  \bottomrule
\end{tabular}
\end{table}

\subsection{DNS over TLS (DoT)}
DNS over TLS (DoT) is specified by RFC 7858~\cite{rfc7858}. According to the specification, the DNS wireformat messages defined in RFC 1035~\cite{rfc1035} are sent over a secure TLS or Datagram TLS connection. IANA reserved port 853/TCP, which should be used by default in all DoT clients and resolvers. Since the packets are sent over a dedicated port, the traffic can be easily recognized, blocked, or filtered by a network administrator. However, it is worth noting that the RFC standard allows using DoT over ports other than 853/TCP; therefore, it is expected that DoT clients and resolvers will have the option to change the port in their configuration.

\subsection{DNS over HTTPS (DoH)}
The DNS over HTTPS (DoH) protocol was adopted as RFC 8484~\cite{rfc8484} in October 2018. Currently, there are two significantly different implementations. The RFC\,8484 compliant approach uses the DNS ``wireformat''~\cite{rfc1035} encapsulated in the HTTPS protocol. The messages are transferred either by secured HTTP GET or POST requests. The second approach uses DNS messages encoded in JSON format described in RFC\,8427~\cite{rfc8427}. The JSON data are then transferred via HTTPS GET. 
Currently, the majority of the DNS providers support both implementations. However, in practice, all DoH enabled browsers and most of the other performance-oriented DoH clients use RFC\,8484 compliant Wireformat messages together with the POST method. Also, browsers and other DoH enabled applications can bypass the DNS resolution in the OS because they communicate with DoH providers directly instead of system calls. Thus, the decision about the usage of DoH is often made by applications, not by OS or users.

Contrary to DoT, DoH does not have a dedicated port. The packets are sent over regular HTTPS (TCP/443). The usage of a well-known port adds a layer of privacy since it is much more difficult for system administrators to block or even recognize DoH from regular HTTPS traffic. Therefore, DoH poses a big challenge for security software and parental control systems since it cannot be inspected or easily blocked. 

The bypass of parental control systems caused the introduction of \textit{secure} DoH providers that claim to block malicious requests. Thus, DoH effectively transfers the visibility and control to DoH providers.

Data centralization is one of the main arguments against DoH usage. Therefore, RFC draft~\cite{pauly-dprive-oblivious-doh} introduced the Oblivious DoH (ODoH) protocol, which separates IP addresses and content by placing a proxy between the user and the DoH resolver. Even though ODoH significantly improves users' privacy, it also impacts the performance due to the added latency by the intermediate proxy.

\subsection{DNS over QUIC}

DNS over QUIC (DoQ), which is currently specified in an active RFC Draft~\cite{huitema-dprive-dnsoquic}, is another form of encrypted DNS security that uses TLS encryption in the form of the QUIC internet protocol. DoQ is very similar to DoT since it encapsulates DNS Wireformat messages (specified in RFC~1035~\cite{rfc1035}) into a QUIC connection. The major difference is the recommendation of port 784/UDP for experimentation.



\section{DoH Resolvers on the Internet}
The use of DoH resolvers might show the acceptance of DoH by the community. We measured the amount of DoH in two ways: (i) by finding the \textit{well-known} DoH providers that are indexed by the community; and (ii) by scanning the whole Internet searching for \textit{unadvertised} working DoH servers.

\subsection{Well-known DoH Providers}
\label{sec:doh-providers}

There are several lists of DoH servers, but none of them is comprehensive. Thus, we aggregated all lists of DoH providers we have found to create our own comprehensive list that consists of 234 well-known DoH providers. These servers are available and verified to work\footnote{The verification was done between April 15th, 2021, and April 23rd, 2021}. The list is published as a new dataset~\cite{MendeleyData2021} on Mendeley Data, and its summary is shown in Table~\ref{tab:well-known-list-stats}.

\begin{table}[t]
\centering
\caption{Summary of our novel comprehensive list of well-known DoH providers}
\begin{tabular}{l c}
\toprule
                          & Number \\
\hline
Total Unique Servers      & 234 \\
Total Unique IPv4 Servers & 131 \\
Total Unique IPv6 Servers & 103 \\
Unique Autonomous Systems & 52  \\
Unique Domain Names       & 110 \\
\bottomrule
\end{tabular}
\label{tab:well-known-list-stats}
\end{table}

\subsection{Finding Unknown DoH Servers on the Internet}
Since the list of well-known providers consists of resolvers owned by organizations, it does not contain small resolvers maintained by individuals that do not want their resolvers to be known. Therefore, we scanned the IPv4 Internet address space to find all the "unknown" resolvers. 

\subsubsection{Scanning Methodology}
To find computers on the Internet with port 443/TCP open we used docker containers on five different machines running scanning software. The steps were: (i) scanning for servers with port 443/TCP open, (ii) finding which ones were DoH resolvers, (iii) confirming the DoH protocol, (iv) identifying the resolvers. 

\paragraph{Scanning for servers with TCP/443 open} The scanning was performed in five computers by uniformly dividing the IPv4 address space. Each machine scanned its address range using the \textit{masscan} tool, which was limited to send 10\,K packets/s. Masscan~\cite{graham2013masscan} is an internet-scale port scanner capable of very high-speed port scanning. 

The details about the scanning can be seen in Table~\ref{tab:scan-internet-amount}. We found 34,078,720 IP addresses with port 443/TCP open. Interestingly, at the time of scanning, Shodan\footnote{\url{https://shodan.io}} listed around 80M IP addresses with port 443/TCP open. Therefore, we performed this scanning two times from different ISPs with a consistent result of 40M IP addresses. At the time of writing, Shodan lists around 40\,million of IP addresses with port 443/TCP open; therefore we believe, that previous Shodan results were inaccurate.

\begin{table}[t]
\centering
\caption{Internet ranges separation for scanning hosts with port 444/TCP open (e.g. Scanner 1: 0.0.0.0-51.0.0.0).}
\begin{tabular}{l c r r}
\toprule
Scanner ID & Range & Host Scanned & Host Found \\\hline
      1  & 0-51  & 855,638,017   &  9,437,184  \\
      2  & 52-103 &  855,638,017 &  8,454,144 \\
      3  & 104-154 &  838,860,801 &  7,405,568 \\
      4  & 155-205 &  838,860,801 &  6,815,744 \\
      5  & 206-255 &  822,083,585  & 1,966,080  \\
      \bottomrule
\end{tabular}
\label{tab:scan-internet-amount}
\end{table}

\paragraph{Finding DoH resolvers} From the open ports we found which ones were DoH servers by creating our own Nmap Lua script. Nmap~\cite{lyon2008nmap} is a well-known multi-functional network scanner that has its own Nmap Script Engine (NSE) for users to develop their own codes. Despite its many scripts, Nmap does not have an official script to find DoH servers.

Therefore, we have developed and published a set of NSE scripts~\footnote{\url{https://github.com/stratosphereips/DoH-Research}} that combines 6 different DoH verification methods. It tests for both HTTP/1 and HTTP/2, and three alternative methods based on GET and POST requests. Since the current Nmap NSE libraries do not support HTTP/2, we had to use external libraries. An example output of the script is shown in Fig.~\ref{fig:dohcheck-nse-example}.

\begin{figure}[t]
    \centering
    \begin{lstlisting}
    PORT    STATE SERVICE
    443/tcp open  https
    | dns-doh-check: 
    |   DoH-JSON: false
    |   DoH-GET: true
    |   DoH-POST: false
    |   DoH2-JSON: false
    |   DoH2-GET: true
    |_  DoH2-POST: false
    \end{lstlisting}
    \caption{Example output of the Nmap NSE script custom developed to find DoH servers }
    \label{fig:dohcheck-nse-example}
\end{figure}

The Nmap script was used to connect to all 40M servers with port 443/TCP and verify which of them had functioning DoH servers. The scan was executed using 40 computers in parallel. Importantly, Nmap was run with \textit{-T 5} ``Insane'' timing mode, and one connection retry attempt.

\paragraph{Resolvers Additional Verification} An additional verification was performed later with a slower Python script to confirm that the resolvers worked, and to measure their properties as reported in Tables~\ref{tab:doh-method-support} and Table~\ref{tab:doh-http-support}.

\paragraph{Identification of Resolvers} The list of resolvers found was further enriched with their reverse DNS record (PTR) and their passive DNS data. When the passive DNS service returned multiple domain names for an IP address, we choose the first result containing the substring ``doh" or ``dns". 

\subsubsection{Scanning Limitations}
The main limitation of our scanning was that it could not find DoH resolvers hosted on infrastructures hosting multiple services behind a single IP address. In such cases, an SNI, or HTTP Host header, or HTTP/2 \textit{:authority} header, is needed for a successful request. Since we did not have the SNI, it was impossible to provide~it. 

\subsubsection{Scan Results}
The DoH scanning found 931 DoH resolvers, which were published as a new dataset for the community~\cite{FoundDoHResolversData}. An analysis of the support of DoH methods can be seen in Table~\ref{tab:doh-method-support}. The majority of the DoH resolvers were compliant with RFC\,8484. The JSON-based approach is only supported by one-third of all resolvers. Additionally, Table~\ref{tab:doh-http-support} shows the support of the HTTP version. Even though RFC\,8484 recommends to use HTTP/2 due to the worse performance of HTTP/1, it can be seen that most resolvers support both versions. 

\begin{table}[t]
\centering
\caption{Supported DoH methods by the resolvers found in our Internet scan.}
\begin{tabular}{l c c}

\toprule
Method & HTTP/1           & HTTP/2         \\ \hline
JSON            & 328 (35.2\,\%)   & 324 (34.8\,\%) \\
GET             & 830 (89.1\,\%)   & 865 (92.9\,\%) \\
POST            & 836 (89.7\,\%)   & 840 (90.2\,\%) \\
JSON, GET       & 328 (35.2\,\%)   & 324 (34.8\,\%) \\
JSON, POST      & 327 (35.1\,\%)   & 322 (34.5\,\%) \\
POST, GET       & 821 (88.1\,\%)   & 819 (87.9\,\%) \\
JSON, GET, POST & 327 (35.1\,\%)   & 322 (34.5\,\%) \\ 
\bottomrule
\end{tabular}
\label{tab:doh-method-support}
\end{table}

\begin{table}[t]
\centering
\caption{Supported HTTP versions by the resolvers found in our Internet scan.}
\begin{tabular}{l c}
\toprule
HTTP Version support & Number of servers  \\
\hline
Only HTTP/1          & 45  (4.9\,\%)   \\
Only HTTP/2          & 86  (9.2\,\%)   \\
HTTP/1 and HTTP/2    & 800 (85.9\,\%)  \\
\bottomrule
\end{tabular}
\label{tab:doh-http-support}
\end{table}

The number of discovered DoH servers is around 4x larger than the number of well-known DoH resolvers currently known by the community. However, multiple resolvers could belong to the same organization and might share the same domain name. Therefore, we performed an analysis of domain names and IP ranges to further describe them.

From the 931 discovered IP addresses, 251 did not have PTR records (which can be considered suspicious). By combining the reverse DNS with the passive DNS data, we obtained domain names for 685 (73.5\%) of the resolvers. To find out how many distinctive resolver providers were found, the hostnames were grouped based on the second-level domains (SLD), finding 131 distinct ones (14\%). The resolvers with the same SLD usually have similar IP addresses. 

For the resolvers without a hostname we group them by their /24 IP range and found 142 distinct IP ranges. The active DoH scanning therefore found DoH resolvers belonging to 273 different organizations, which is twice more than the current number of the different organizations contained in the existing lists of well-known DoH providers. Additionally, we only found 32 of the well-known DoH IP addresses among the new list of DoH servers, which is expected since large providers might require SNI. Table~\ref{tab:hostname-analysis} summarizes these findings.

\begin{table}[t]
\centering
\caption{Features of IP addresses of new DoH resolvers found by the scan.}
\begin{tabular}{lc}
\toprule
Feature & Amount\\
\hline
Total number of unique IP addresses       & 931 (100\,\%) \\
IP addresses with PTR records         & 685 (73.5\,\%) \\
IP addresses without PTR records      & 251 (26.9\,\%) \\
Unique SLD                            & 131 (14\,\%)\\
Unique /24 prefixes                   & 142 (15.2\,\%) \\
Assumed \# of unique providers        & 273 (29.3\,\%) \\
Number of discovered well-known resolvers & 32  (3.4\,\%)\\
Number of discovered ``unknown" resolvers & 899  (96.4\,\%) \\
\bottomrule
\end{tabular}
\label{tab:hostname-analysis}
\end{table}

\section{Trends in the Adoption of Encrypted DNS traffic}
Since the adoption of the DoH and DoT standards, many changes came upon the applications and tools used to resolve DNS. An understanding of the trends of adoption may give researchers an estimation of the volume of traffic that is encrypted and how many users are transitioning into the technology. 

This section first describes the networks and organizations where the traffic comes from, and then it analyzes the traffic from four perspectives: (i) amount and comparison of encrypted DNS traffic in each organization, (ii) analysis of the stationarity and trends of the traffic in each organization, (iii) ratios of relationship between traffic, with their stationarity.

\subsection{Description of the Source Networks and Datasets}
\label{sec:desc-organizations}
The measurement of encrypted DNS protocols was done through a collaboration between three major organizations. The first organization is a large European ISP provider, the second organization is a large European University, and the third organization is a global security company.

\paragraph{Large European ISP Provider (Organization 1)}
Organization 1 is a national research and education network in an EU Country that interconnects many academic institutions, research organizations, governmental offices, and others. It represents around 500,000 users. Organization 1 develops high-speed network traffic monitoring and analysis tools, which are deployed on multiple 100\,Gb/s links operated at the perimeter of the infrastructure. There are six monitoring probes for the perimeter, equipped with special hardware cards with FPGA to accelerate the process. The network monitoring probes are set with 5 minutes of active and 30 seconds of passive timeout to produce IPFIX flow records extended with custom information fields. The flow collector receives an average rate of about 150\,K flow records per second from 8 peering lines.

Since Organization 1 is a large ISP, it can see communication coming from the connected institutions and also transit traffic. The transitive traffic might generate duplicate flows since the communication is passing by two monitoring probes. However, the measurement infrastructure can deal with duplicates and filters them out. The measurement of IPv4 and IPv6 traffic is performed on a flow collector by filtering and storing IP Flow data in the following way:

\begin{itemize}
\item \textbf{DoH} traffic is filtered from IP Flows extended with TLS handshake information. Organization 1 counts only \textit{1-RTT} TLS handshakes; \textit{0-RTT} handshakes are not counted. DoH traffic is therefore measured as the amount of Client Hello packets with SNIs of well-known DoH providers transferred via port 443/TCP. Even though we composed a comprehensive list of well-known providers in Section~\ref{sec:doh-providers}, it is not feasible for Organization 1 to create a complete filter with such a high number of SNIs. Therefore, we reduced the whole list to only the ten most suitable DoH providers. The list of the providers used for DoH measurement is shown in Table~\ref{tab:WellKnownDoHResolvers}.

\item \textbf{DNS over TLS} traffic was also filtered from IP Flows extended with TLS information, which ensures the occurrence of Client Hello packets. However, in this case, only port filtering was used. Therefore, DoT traffic is calculated as the number of Client Hello packets observed on port 853/TCP.

\item \textbf{DNS over QUIC} traffic was counted from traditional IP Flow data. Since QUIC obfuscates even the handshake packets by encryption, it is unfeasible for Organization 1 to perform DPI for QUIC recognition due to the lack of available computational power. Therefore, DoQ traffic is calculated as the number of flows observed on port 784/UDP.

\item \textbf{DNS} traffic was measured from traditional IP Flow data by filtering by the port number of 53/UDP and 53/TCP.

\item \textbf{Total} traffic was measured from traditional IP Flow data without any filtering. 
\end{itemize}

\begin{table*}[t]
\small
\caption{Major DoH Providers}
\begin{center}
\begin{tabular}{l c c c c c}
\toprule
\textbf{} & \makecell{\textbf{IPv4-1}} & \makecell{\textbf{IPv4-2}}  & \makecell{\textbf{IPv6-1}} & \makecell{\textbf{IPv6-2}} \\
\hline
dns.cloudflare.com & 1.1.1.1 & 1.0.0.1 & 2606:4700:4700::1111 & 2606:4700:4700::1001 \\
dns.google.com & 8.8.4.4 & 8.8.8.8 & 2001:4860:4860::8888 & 2001:4860:4860::8844 \\
dns.aa.net.uk & 90.155.62.13 & 90.155.62.14 & N/A & N/A \\
dns-nyc.aaflalo.me & 104.27.159.50 & 104.27.158.50 & N/A & N/A \\
dns.adguard.com & 104.20.31.130 & 104.20.30.130 & 2a10:50c0::ad1:ff & 2a10:50c0::ad2:ff \\
doh.cleanbrowsing.org & 192.124.249.8 & N/A & N/A & N/A \\ 
nic.cz & 193.17.47.1 & 185.43.135.1 & N/A & N/A \\
dns.nextdns.io & 104.31.88.168 & 104.31.89.168 & 2a07:a8c0::1c:7db6 & 2a07:a8c1::1c:7db6 \\
dns.brahma.world & 104.27.170.14 & 104.27.171.14 & N/A & N/A \\
dns1.dnscrypt.ca & 69.165.220.221 & N/A & 2620:fe::fe & 2620:fe::9 \\
libredns.gr & 116.202.176.26 & N/A & N/A & N/A \\
\bottomrule
\end{tabular}
\label{tab:WellKnownDoHResolvers}
\end{center}
\end{table*}

\paragraph{Large European University (Organization 2)}
Organization 2 is a large European University. The traffic is being captured in an egress network connection that includes a complete faculty with approximately 2,200 computers. This network includes mostly cable-connected devices and some WiFi devices, but not the Eduroam WiFi network; and therefore the large majority of computers are desktop computers and servers.

Organization 2 uses two redundant methods to capture traffic, one using the Zeek Security Monitor~\cite{Zeek}, and other using the Argus network sensor~\cite{Argus}.

All the captured traffic was filtered so only flows \textit{started} by computers in the IP range of Organization 2 were included (both IPv4 and IPv6). All data from Organization 2 includes this filter.

The data from Organization 2 has the following features:
\begin{itemize}
    \item Number of flows: Total flows generated by the organization. From Zeek conn.log.
    \item Number of flows to port 443/TCP: Regardless of the state. From Zeek conn.log.
    \item Number of flows TLS Established: Established TLS flows in any port. From Zeek ssl.log
    \item Number of DoQ flows: Flows to port 784/UDP. From Zeek conn.log.
    \item Number of DoT flows: Flows to port 853/TCP. From Zeek conn.log.
    \item Number of DoH flows: Flows to well-known DoH providers~\ref{sec:doh-providers}. From Zeek ssl.log.
    \item Number of DNS flows: Flows going out to port 53/UDP, but not to the main DNS server of the organization. From Zeek conn.log.
    \item Number of unique source IPs: Count of unique IPv4 and IPv6 IPs. From Zeek conn.log.
\end{itemize}

The DNS flows were filtered to discard flows going to the official DNS server of the organization because (i) the traffic from the local network to that server does not appear in the capture, and (ii) we discard spurious connections from the Internet to the main DNS server.

The DoH flows were computed by first filtering all established TLS traffic, i.e. to port 443/TCP with an established TLS session, and then by filtering the IP address of the DoH providers shown in Table~\ref{tab:WellKnownDoHResolvers}.

\paragraph{Global Security Company (Organization 3)}
Organization 3 is a large, global security company protecting hundreds of millions of endpoints. To protect clients against security issues, the traffic produced in the endpoints is analyzed and blocked in the endpoints. The solution processes HTTP and HTTPS traffic from both IPv4 and IPv6. It uses URL detection algorithms to analyze and protect against threats as well as a full content filtering to stop malware. As a result, over 300 billion URLs are checked each month.

As part of the research to identify new and emerging threats, the threat intelligence gathered from the endpoints can be queried to produce trends and statistics, for instance, to count the number of DoH and DNS requests.

For privacy reasons, Organization 3 can not share the absolute amounts for their measurements, and therefore all the analysis are done by ratios of values, such as \textit{DoH flows per user}.

\subsection{Amount of Encrypted DNS Traffic}
\label{sec:amount-encrypted-dns-traffic}
The first descriptive analysis is done on the amount of traffic of every encrypted protocol that has been seen by the three organizations. First, we show the total amounts for Organization 1 and 2, and then we compare the combinations of ratios of values between all three organizations.

The DoH traffic was filtered in all organizations using the list of DoH resolvers in Table~\ref{tab:WellKnownDoHResolvers} instead of our new comprehensive list because the short list contains the providers that cover the majority of the traffic, and because Organization 1 has a technical limit in their filtering capabilities.

\begin{figure*}[ht]
    \centering
    \includegraphics[width=\linewidth]{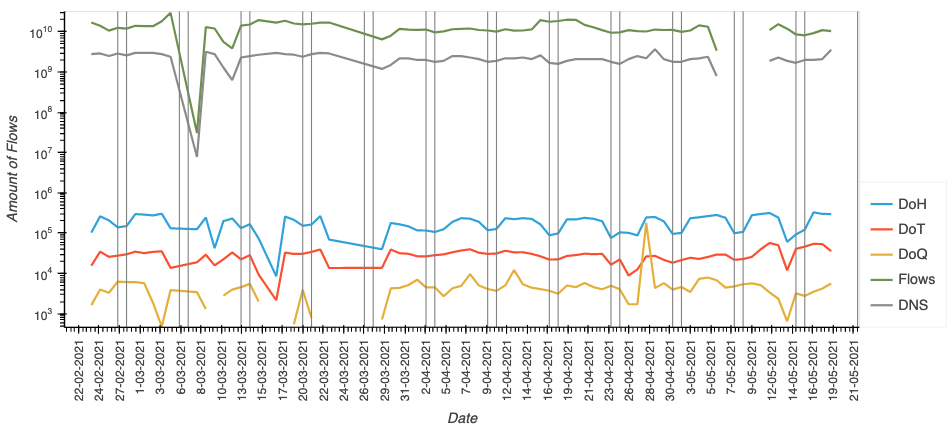}
    \caption{Amount of flows for Organization 1 in logarithmic scale. Total flows, amount of DNS flows, DoH flows, DoQ flows, and DoT flows. From February 24th 2021 to May 15th 2021. Vertical lines are weekends.}
    \label{fig:org1-totals}
\end{figure*}

The traffic of Organization 1, described in Subsection~\ref{sec:desc-organizations}, is shown in Figure~\ref{fig:org1-totals}. This Figure uses a logarithmic $y$ axis and computes the values per day. The amount of DoH traffic per day has \textit{mean=}181,794 and \textit{STD=}78,331. The DoT traffic per day has a mean of 28,395 and a \textit{STD=}10,120; being in average 6.4x times smaller than DoH. Note that the amount of DoH traffic is larger than the amount of DoT by one order of magnitude. The DoQ traffic per day has a \textit{mean=}6,235 and a \textit{STD=}20,131. The amount of DoQ traffic is approximately 29 times smaller than DoH, but still significant.

Regarding the percentages of each protocol for Organization 1, the percentage of DoT traffic to the total flows had a \textit{mean=}0.00001\%, and \textit{STD=}0.00007. The percentage of DoH traffic to the total flows had a \textit{mean=}0.00007\% and \textit{STD=}0.0004.

For the use of any encrypted DNS protocol in comparison with all DNS-related protocols, Organization 1 has, on average, a 0.01\% of its DNS traffic being encrypted.

\begin{figure*}[ht]
  \centering
  \includegraphics[width=\linewidth]{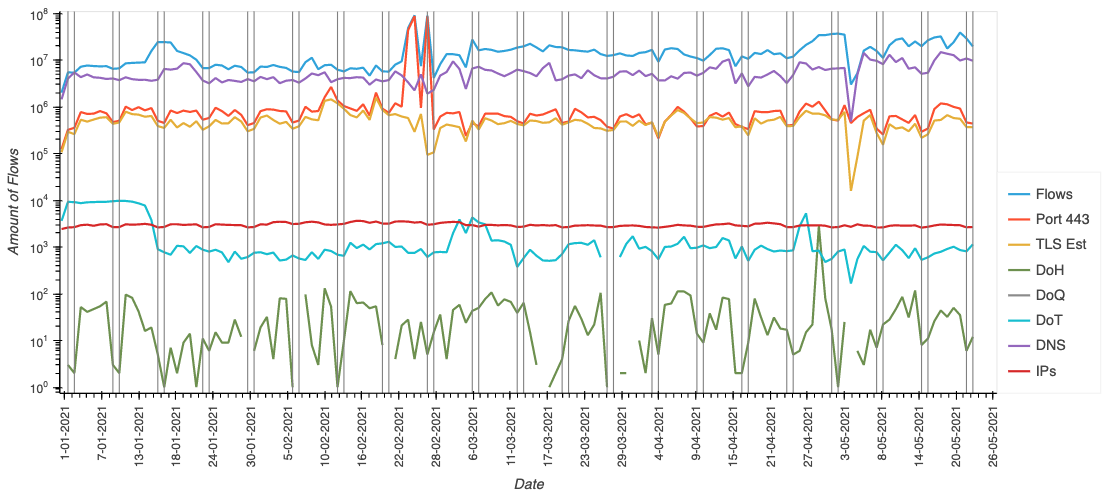}
  \caption{Traffic per day in logarithmic scale from Organization 2 showing total amount of flows, amount of DoH flows, amount of DNS flows, and amount of established TLS flows, amount of flows to port 443/TCP, amount of DoT flows, amount of DoQ flows, amount of DNS flows and amount of IP addresses. From January 1st, 2021 to May 23th, 2021. Vertical lines are weekends.}
  \label{fig:org2-totals}
\end{figure*}

The traffic of Organization 2, described in Subsection~\ref{sec:desc-organizations}, is shown in Figure~\ref{fig:org2-totals}. This Figure uses a logarithmic $y$ axis and computes the values per day. In the Figure it can be seen that the DoT traffic is larger than the DoH traffic, which is opposite to Organization 1. We could not reconcile this difference and we attribute it to the potentially larger number of Android phones in Organization 1. The amount of DoH traffic is in average 35 times smaller than DoT. The amount of DoQ traffic is almost continually zero. The mean and standard deviation values for all the measurements are shown in Table~\ref{tab:org2:means}. The STD of the amount of flows is large due to the infection happening around Feb 28th and the outage happening around May 3rd.

Figure~\ref{fig:org2-totals} shows that the DoT traffic had an important decrease around mid January 2021, that is not an anomaly but a consequence of some unknown change in the process. The reasons for this strong decline could not be found, but as shown in Subsection~\ref{sec:past-trends} a similar but opposite phenomenon happened in 2020. Around February 22th there was a peak in the amount of flows to port 443/TCP (top red line). This peak corresponds to an infected computer inside the organization that scanned ports 443/TCP on the Internet. However, it can be seen that the TLS Established traffic (yellow line) does not have that peak because most connections were not established. 

\begin{table}[t]
\centering
\caption{Mean and STD of the main flow amounts for Organization 2 from January 1st, 2021 to May 23th, 2021. Mean is amount of flows for all values except for IPs that it is amount of unique IPs.}
\begin{tabular}{l r r}
\toprule
 Value & Mean &  STD \\
\midrule
DoH & 50.3 & 237.6 \\
DoT & 1782.4 & 2459.0 \\
DoQ & 0 & 0.1 \\
DNS & 5,524,360.2 & 2,535,076.7 \\
Flows & 15,504,487 & 12,290,761.5 \\
TLS & 506,300.9 & 222,479.8 \\
Port443 & 2,227,208.6 & 10,780,265.3 \\
IPs & 2,971.4 & 253.6 \\
\bottomrule
\end{tabular}
\label{tab:org2:means}
\end{table}

It can also be seen in Figure~\ref{fig:org2-totals} that the number of unique IP addresses (both IPv4 and IPv6) is quite constant (red line in the middle), with a small standard deviation of 253. Around May 3rd there was a small network outage that reduced the amount of traffic collected and added some bias to the measurements.

Regarding the percentages of each protocol for Organization 2, the percentage of DoT traffic to the total flows had a \textit{mean=}0.0002\,\%, and \textit{STD=}0.0003. Compared with Organization 1, the percentage of DoT traffic is 20 times larger in Organization 2 despite its total amount being 16 times larger in Organization 1. 

The percentage of DoH traffic to the total flows for Organization 2 had a \textit{mean=}0.000003\,\% and \textit{STD=}0.000007. The percentage of DoH traffic is 23x times larger in Organization 1 than in Organization 2, and its total amount is 3.6 times larger in Organization 1.

Regarding the use of any encrypted DNS protocol in comparison with all DNS-related protocols, Organization 2 has, on average, a 0.033\% of its DNS traffic being encrypted. This is 3 times more than Organization 1.

The traffic of Organization 3, described in Subsection~\ref{sec:desc-organizations}, is shown as ratios of other values. The first comparison is on the amount of DoH flows per unit user per country per day, normalized by country population. The countries used are the top 8 countries with the largest amount of DoH flows in total. This ratio is computed by dividing the DoH traffic for the whole country per day, by the total number of unique users sending DoH in all countries, and again dividing by the population of the country. 

\begin{figure*}[ht]
  \centering
  \includegraphics[width=0.8\linewidth]{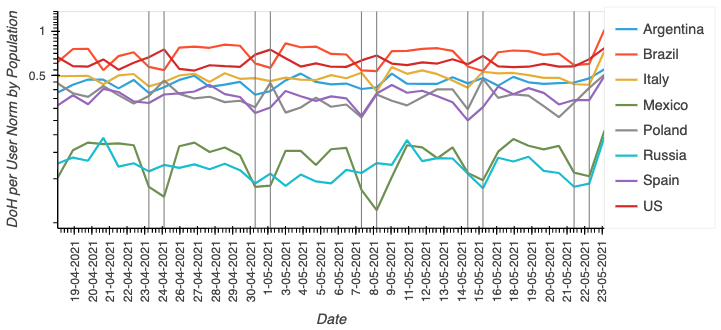}
  \caption{Organization 3 DoH traffic per user per country, normalized by the relative size of the population of each country. Vertical lines are weekends. Population estimation in millions taken from datacommons.org}
  \label{fig:org3-doh-country-user-norm}
\end{figure*}

Figure~\ref{fig:org3-doh-country-user-norm} shows the DoH values per user per country normalized by the relative population of the country. Population data taken from Datacommons\footnote{\url{https://datacommons.org/}}: Argentina 44.94\,M, Brazil 211\,M, Italy 60.36\,M, Mexico 127.6\,M, Poland 37.97\,M, Russia 144.4\,M, Spain 46.94\,M, US 328.2\,M. The normalized comparison shows that even though US and Brazil are still the top senders, they are not by a large margin, and Italy appears like a strong DoH user followed by Argentina. The exact mean and standard deviation values are shown in Table~\ref{tab:org3-mean-std-doh-user}.

We did not normalize the values with the number of users that Organization 3 has in each country, but we assume the population follows the size of the countries since Organization 3 is global.

\begin{table}[t]
    \caption{Mean and Standard Deviation for the top 8 countries by DoH count, of ``DoH flow count-per-user, per day, per capita'', in Organization 3.}
    \label{tab:org3-mean-std-doh-user}
    \begin{tabular}{l r r}
    \toprule
    Country & Mean & STD \\
    \midrule
    Argentina & 0.44 & 0.03 \\
    Brazil & 0.69 & 0.1 \\
    Italy & 0.49 & 0.05 \\
    Mexico & 0.14 & 0.03 \\
    Poland & 0.35 & 0.05 \\
    Russia & 0.12 & 0.02 \\
    Spain & 0.35 & 0.04 \\
    US & 0.61 & 0.05 \\
    \bottomrule
    \end{tabular}
\end{table}


\paragraph{Stationarity of Values}
To better understand if the traffic is growing, decreasing, or having any trend, we tested the stationarity of the measurements. It was tested using the Augmented Dickey–Fuller test (ADF)~\cite{AugmentedDF}, which is the standard test of non-stationarity of time series. A stationary time series is one whose properties do not depend on the time at which the series is observed~\cite{hyndman2018forecasting}. Time series that show trends or seasonality are non-stationary. The implementation used was the function \textit{adfuller()} from the Python library \textit{statsmodels.tsa.stattools}.

The results of the ADF test on the raw amount of data for the three organizations can be seen in Table~\ref{tab:adf-values-orgs}. The test shows that in Organization 1 the only Stationary value is the DoT traffic. The rest of the values seem to have a trend or a strong seasonality. For Organization 2 most of the values are stationary, except the number of IP addresses that according to its linear model it seems to be growing. Note that the DoT traffic of Organization 2 is ignored since there were only two days with a value of 1.

In particular for the amount of IP addresses of Organization 2, the ADF test failed to reject the null hypothesis that the series is non-stationary with a confidence of 95\,\%, and therefore we conclude that the number of IPs is non-stationary. This result may seem contradictory with Figure~\ref{fig:org2-totals} where the line for IP addresses seems \textit{very} stationary. However, a detailed study of the time series of IP addresses shows that it has a strong cycles due to weekends. The result of non-stationarity is therefore correct because according to the theory, a time series of cyclic behavior (without trends or seasonality) is stationary \textit{only if the cycles are not of fixed length}. Since the cycles have a fixed length due to the weekends, the series is therefore non-stationary~\cite{hyndman2018forecasting}.

\begin{table}[t]
  \caption{Augmented Dickey-Fuller test for Stationarity of values in the three organizations for the duration of their measurements in 2021. Confidence interval of 95\,\%.}
  \label{tab:adf-values-orgs}
    \begin{tabular}{llrrl}
    \toprule
    Org & Value & ADF Stat & p-value & Conclusion \\
    \midrule
    1 & DoH & -2.7 & 5.95e-02 & Non-Stationary \\
    1 & DoT & -1.1 & 6.91e-01 & Non-Stationary \\
    1 & DoQ & -8.7 & 2.66e-14 & \textbf{Stationary} \\
    1 & Flows & -2.03 & 2.70e-01 & Non-Stationary \\
    1 & DNS & -4.3 & 3.01e-04 & \textbf{Stationary} \\
    2 & DoH & -11.6 & 1.77e-21 & \textbf{Stationary} \\
    2 & DoT & -2.2 & 1.80e-01 & Non-Stationary \\
    2 & DoQ & -12.09 & 2.13e-22 & \textbf{Stationary} \\
    2 & DNS & -5.3 & 5.32e-06 & \textbf{Stationary} \\
    2 & Flows & -4.6 & 1.27e-04 & \textbf{Stationary} \\
    2 & TLS & -6.6 & 4.42e-09 & \textbf{Stationary} \\
    2 & Port443 & -3.3 & 1.33e-02 & \textbf{Stationary} \\
    2 & IPs & -2.07 & 2.56e-01 & Non-Stationary \\
    3 & DoH & -0.1 & 9.44e-01 & Non-Stationary \\
  \bottomrule
    \end{tabular}
\end{table}

\begin{figure}[ht]
  \centering
  \includegraphics[width=\linewidth]{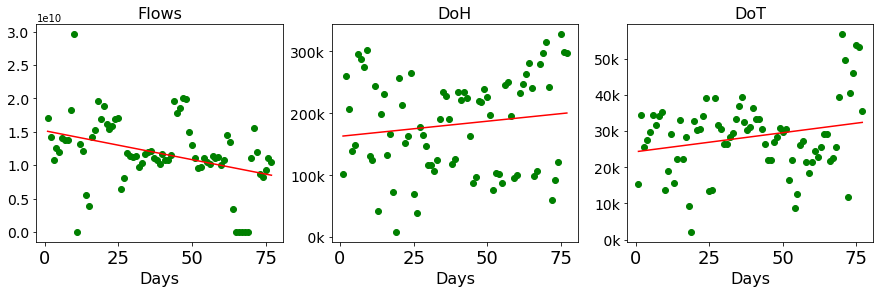}
  \caption{Linear regression of non-stationary values on Organization 1. The x-axis is in numbers of days since Feb 24th, 2021. The y-axis is amounts.}
  \label{fig:lin-reg-values-org1}
\end{figure}

\paragraph{Linear Regression of Trends}
The last analysis for the raw values (not ratios) is a linear regression on the values to evaluate if the linear coefficient is positive or negative and its overall trend. This analysis evaluates the trend for the values that were found not to be stationary. Figure~\ref{fig:lin-reg-values-org1} describes the trends, showing that the amount of flows is probably decreasing, the amount of DoH flows slowly increases, and the amount of DoT flows is slowly increasing too. 

\begin{figure}[ht]
  \centering
  \includegraphics[width=\linewidth]{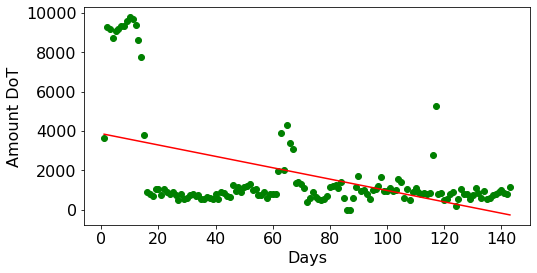}
  \caption{Linear regression of non-stationary DoT values on Organization 2. The trend seems to be decreasing. The x-axis is in numbers of days since January 1st, 2021}
  \label{fig:lin-reg-value-dot-org2}
\end{figure}

These results for Organization 1 are different from Organization 2. Figure~\ref{fig:lin-reg-value-dot-org2} shows how for Organization 2 the DoT trend is decreasing.

\begin{figure}[ht]
  \centering
  \includegraphics[width=\linewidth]{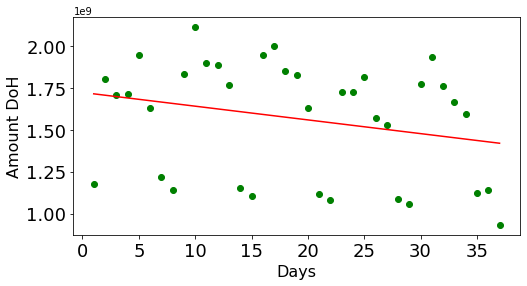}
  \caption{Linear regression of non-stationary DoH values on Organization 3. The trend seems to be decreasing. The x-axis is in numbers of days since April 18th, 2021}
  \label{fig:lin-reg-value-doh-org3}
\end{figure}

The most important comparison may be between the DoH trend of Organization 3, shown in Figure~\ref{fig:lin-reg-value-doh-org3}, which seems to be slightly decreasing and Organization 1 which seems to be slightly increasing (Figure~\ref{fig:lin-reg-values-org1}). 

\subsection{Ratios Between Organizations}
\label{sec:ratios-organizations}

\paragraph{Ratio DoH per 1 million flows}
Given the difference amount of users on each organization it was necessary to compute a ratio-per-user and a ratio-per-amount-of-flows in order to compare the changes in the traffic. These ratios are based on the idea that given large enough networks, the users tend to generate a similar amount of traffic. Therefore, computing the ratios-per-user gives an idea that is independent of the amount of users.

The first ratio comparison done is the \textit{number of DoH flows per 1 million flows}, computed each day. This ratio gives an idea of how much DoH traffic is generated in comparison with the total traffic, but independently of the amount of users seen on each organization. Instead of comparing with the total amount of flows, which is a varying quantity, comparing with the DoH flows per 1 million flows is a much more stable and \textit{transferable} metric. 

\begin{figure}[ht]
  \centering
  \includegraphics[width=\linewidth]{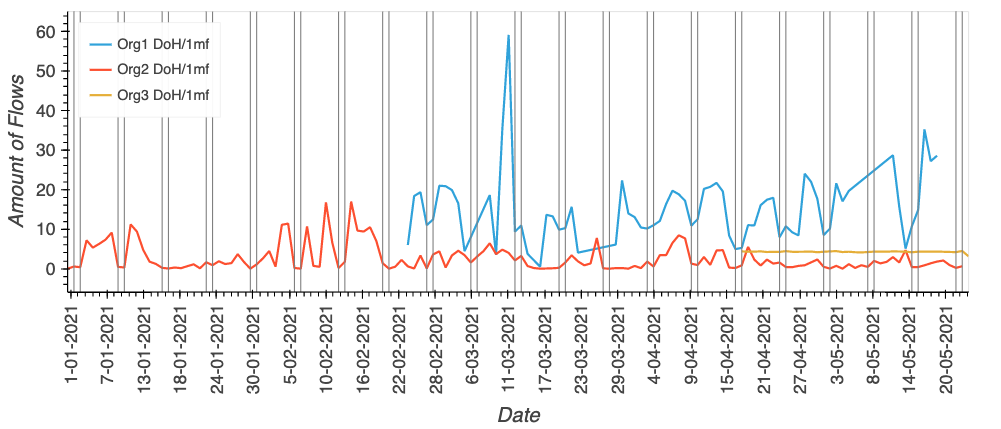}
  \caption{Comparison of the \textit{amount of DoH flows per 1 million flows} for all Organizations. The ratios were processed to discard anomalies.}
  \label{fig:ratio-doh-flows-org1-org2-org33}
\end{figure}

The comparison of ratios for all three organization is shown in Figure~\ref{fig:ratio-doh-flows-org1-org2-org33}. Because each organization had access to a slightly different time-frame, the Figure shows an overlap only in the last 40 days. More importantly, the comparison shows that the ratio of DoH flows per 1 million flows for all three organizations is \textit{comparable} and not with large differences. For Organization 1 (top blue line) the mean is 15.2 (STD=8.9), for Organization 2 (bottom red line) the mean is 3.2 (STD 7.4) and for Organization 3 (middle yellow line) the mean is 4.25 (STD 0.2). Even though the data does not come from the same exact distribution according to an ANOVA test, the values are a good indicator and estimator of this ratio for other networks.

The second evaluation of the ratio of DoH per 1 million flow was regarding its stationarity, and it used again the Augmented Dickey–Fuller test (ADF) (described in Subsection~\ref{sec:amount-encrypted-dns-traffic}). The results of the test show that for Organization 1 the ratio had a statistic of -3.62, with a p-value of 5.35e-03, and therefore using a confidence interval of 95\,\% we can estimate that it is Stationary. For Organization 2, the ratio had a statistic of -11.28, with a p-value of 1.40e-20, and therefore using a confidence interval of 95\,\% we can estimate that it is Stationary. For Organization 3 the ratio had a statistic of -4.59, with a p-value of 1.34e-04 and therefore using a confidence interval of 95\,\% we can estimate that it is Stationary.

\paragraph{Ratio DoH per 1 million DNS flows}

\begin{figure}[ht]
  \centering
  \includegraphics[width=\linewidth]{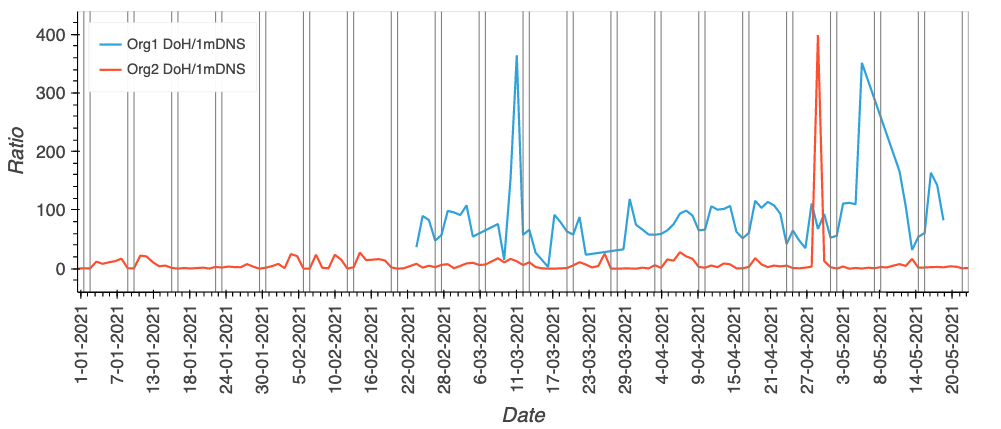}
  \caption{Comparison of the \textit{amount of DoH flows per 1 million DNS flows} for Organization 1, and 2. The ratios were processed to discard anomalies.}
  \label{fig:ratio-doh-dns-org1-org2}
\end{figure}

The second ratio comparison done is regarding the amount of DoH flows per 1~million DNS flows. This ratio, similar to the last one, allows a comparison between the organizations regardless of the amount of users on them. In this case we compare DoH with DNS, since both protocols are intimately related and we may even expect a small decrease in DNS if DoH grows. Figure~\ref{fig:ratio-doh-dns-org1-org2} shows the ratio comparison. Similarly as the previous ratio, even though the data does not come from the same distribution, the numbers are \textit{comparable} and with some similarity. For Organization 1 the mean of the ratio is 86.5 (STD 56.9) and for Organization 2 the mean of the ratio is 8.88 (STD 33.5). 

Regarding their stationarity, using the ADF test, for Organization 1 the statistic is -3.6, with a p-value of 5.57e-03, therefore using a confidence interval of 95\,\% we can conclude that its ratio is Stationary. For Organization 2 the statistic is -11.6, with a p-value of 2.43e-21, therefore using a confidence interval of 95\,\% we can conclude that the ratio is Stationary.

\subsection{Past trends in Encrypted DNS}
\label{sec:past-trends}

\begin{figure*}[ht]
  \centering
  \includegraphics[width=\linewidth]{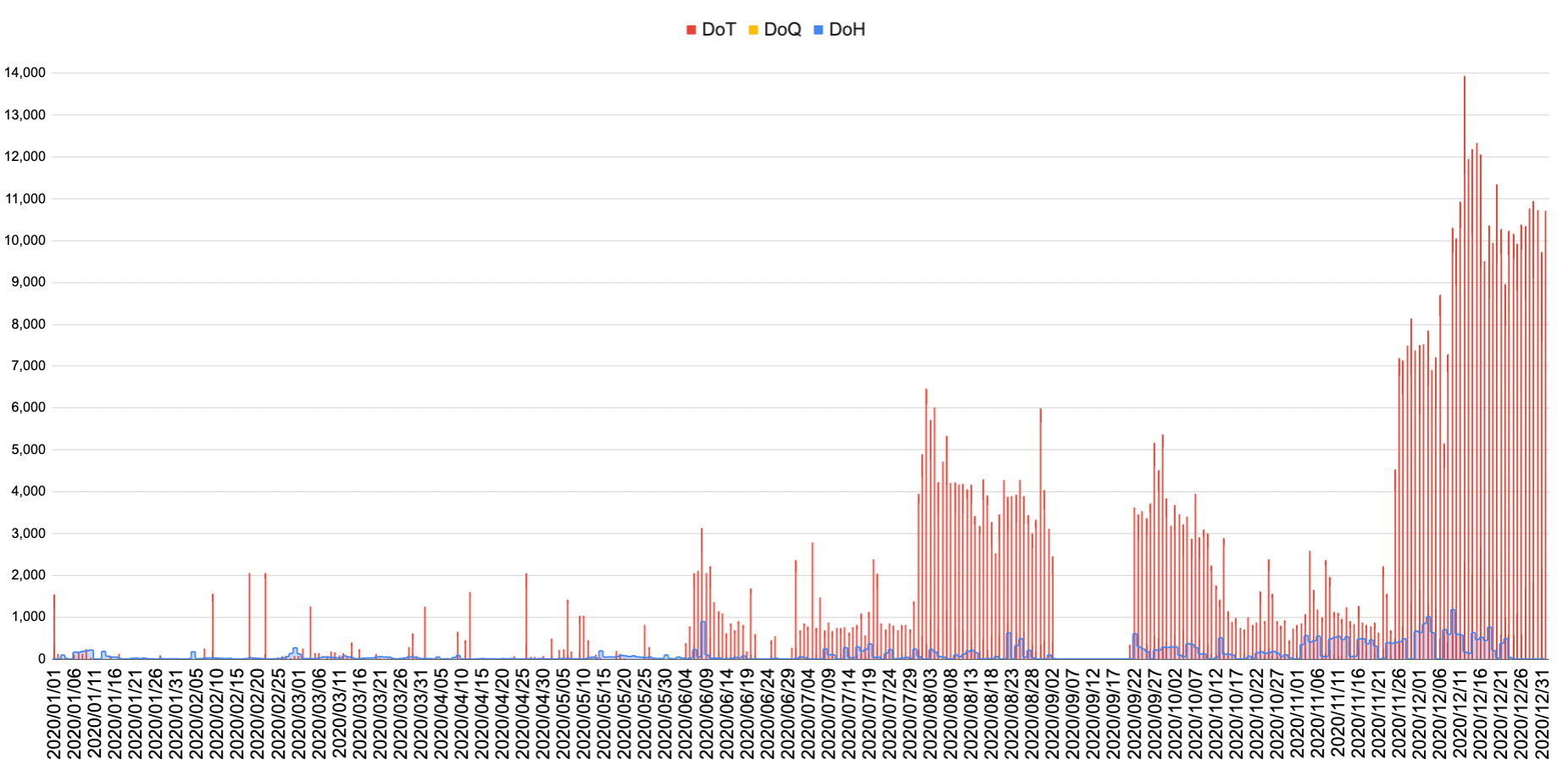}
  \caption{Amount of DoH/DoT/DoQ requests during 2020 in the traffic of Organization 2.}
  \label{fig:org2-2020-doh-dot-doq}
\end{figure*}

Given that almost five months of traffic may not be sufficient to see larger trends, we accessed past traffic captured by Organization 2 during 2020 using Argus sensors. This traffic was not used for comparison with the other organizations since they did not retain old traffic. This traffic is a measurement of DoH, DoT and DoQ flow protocols, as shown in Figure~\ref{fig:org2-2020-doh-dot-doq}. Apart from a sensor problem that broke the capture from 2020/09/09 to 2020/09/23, it can be seen that there are two clear growing trends for DoT and DoQ. In particular from 2020/08/01 the DoT traffic grew 3x in a couple of days. The DoQ traffic continued to be around 4,200 request per day from August 1st to October 10th, after which it lowered to 1,100 requests per day until November 24th. On November 25th it started to grew, reaching a peak on December 12th of 14,000 requests per day. Given that the amount of IPs in Organization 2 remained almost constant, it can be seen that the \textit{peak} ratio of DoT per IP address was around 5.6 requests per day. 

This is a large change if it is compared to the mean of the ratio DoH/IP address for 2021, which was 0.016. The peak ratio of the DoT per IP in 2021 was 350x larger than the mean value for the five first months of 2021.

The growing trend in the total amount of encrypted DNS traffic can better be seen in Figure~\ref{fig:teaser}, which shows the sum of DoH, DoQ and DoT in Organization 2. The Figure uses one line per month and each value corresponds to the same day in the month. It is clear then than the first months of 2021 had more encrypted DNS traffic than all 2020.

\section{Discussion}

\paragraph{DoH Trends} DoH traffic has grown from 2020 to 2021 according to our measurements, but in the first five months of 2021 its growing ratio has become stationary. The percentage of \textit{DoH to total flows} is larger in the ISP provider (Organization 2) than in the university (Organization 2). These results are consistent with previous work~\cite{lu2019}. The difference of \textit{DoH flows per 1 million flows} and \textit{DoH flows per 1 million DNS flows} is significantly similar in the three organizations with a near constant mean, which supports the idea that DoH has become stationary in several locations around the world. The only apparent grow in DoH is in the ISP Organization 1. 

For the global security company (Organization 3) DoH has gone from stationary to slightly decreasing. However, when DoH was measured per user, by country, and normalized by population, the amount of DoH was stable. US and Brazil are the countries with larger \textit{DoH-per-user-per-capita} but Italy and Argentina are close behind.

\paragraph{DoT Trends} DoT traffic seems to be much larger in the ISP (Organization 2) than in the other organizations. Showing a non-stationary growth in this organization. However, it shows an actually decrease in Organization 2 on mid-January 2021. The absolute number of DoT flows is larger than DoH in all the traffic captures combined.

\paragraph{DoQ Trend}
The amount of DoQ traffic was only significant in Organization 1, with a small but stationary amount; almost zero in Organization 2; and not measured in Organization 3. In may be possible that the differences are due to artifacts in the network capture mechanisms. 

\paragraph{DoH Resolvers}
There are 234 current DoH resolvers that are well-known by the community. However, we found 931 working DoH servers that were \textit{not published}. These unpublished DoH resolvers are expected to be used soon as part of new service offerings. From these unpublished resolvers, 73\,\% do not have a reverse DNS hostname, which is considered suspicious. Although most of them implement HTTP/1 and HTTP/2 protocols, there is a 4.9\,\% that only implements HTTP/1, which is not recommended. The reason why these resolvers only use HTTP/1 is unclear. From these 931 unpublished resolvers, only 3\% are in our list of well-known providers.

These results show that DoH cannot be blocked by filtering lists of well-known DoH resolvers because there are 4 times more resolvers working in the Internet. This limitation may play an essential role in the development of future malware and exfiltration tools. 

More importantly, Organization 1 has already seen in its traffic five IP addresses of the 931 unknown IP resolvers being used, corresponding to the passive DNS hostnames \textit{cache2.alidns.com}, \textit{doh.li}, and \textit{security.cloudflare-dns.com}.

\paragraph{Comparison between Organization}
Despite the differences in size, type of organization and scope, it is possible to say that the \textit{ratio} of DoH traffic growth in the beginning of 2021 is similar and almost constant for all organizations. Organization 1 is the most different, showing a small growth in the DoH ratio for May 2021.

\section{Conclusion}
Since the introduction of encrypted DNS proposals, there has not been a comprehensive analysis of the adoption and trends of the protocols. This paper presents an analysis of the traffic on DoH, DoT and DoQ protocols in three large organizations, together with a Internet scan of unpublished DoH resolvers.

DoT traffic seems to be growing in some organizations and has a large volume of traffic considering all absolute numbers. It probably produces more global traffic than DoH.

DoH traffic increased during 2020, but it has been stationary in the first five months of 2021. Given that we found 4 times more DoH resolvers on the Internet than are well-known, we expect that this grey-zone of DoH providers will manifest soon in network traffic in relation to official services and security abuses. Additionally, the relatively high amount of grey-zoned DoH providers proves, that DoH cannot be blocked by using list of providers.

The global trend of DoH traffic in all three organizations is comparable and stationary after a strong increase in 2020. Some countries seem to have larger traffic of DoH per capita, with US and Brazil leading the charts and Italy and Argentina behind.

Future work will include a longitudinal comparison of the traffic after some months to evaluate the trends. It will also include an analysis of unpublished DoT servers.

\begin{acks}
Authors would like to thank Veronica Valeros from the Stratosphere Laboratory for her support in producing the dataset and editing the text. This work was partially funded by Avast and also received funding from the European Union’s Horizon 2020 research and innovation programme under grant agreement No 833418
\end{acks}

\bibliographystyle{ACM-Reference-Format}
\bibliography{references.bib}

\newpage

\section{Appendices}

\subsection{Sources of DoH servers}
The following list of URLs were used as source to create our comprehensive dataset of well-known DoH providers.

\begin{itemize}
    \item \url{https://developers.google.com/speed/public-dns/docs/doh/json}
    \item \url{https://blog.nightly.mozilla.org/2018/06/01/improving-dns-privacy-in-firefox/}
    \item \url{https://github.com/curl/curl/wiki/DNS-over-HTTPS}
    \item \url{https://help.keenetic.com/hc/en-us/articles/360007687159-DNS-over-TLS-and-DNS-over-HTTPS-proxy-servers-for-DNS-requests-encryption}
    \item \url{https://dnsprivacy.org/wiki/display/DP/DNS+Privacy+Public+Resolvers}
    \item \url{https://kb.adguard.com/en/general/dns-providers}
    \item \url{https://applied-privacy.net/services/dns/}
    \item \url{https://www.pacnog.org/pacnog24/presentations/DoT-DoH-DNS-Privacy.pdf}
    \item \url{https://www.privacytools.io/providers/dns/}
\end{itemize}

\subsection{Masscan Configuration}

An example of the masscan command used for the first IP range is: 

\begin{Verbatim}[fontsize=\footnotesize]
masscan -p 443 0.0.0.0-51.0.0.0/0 --exclude 
255.255.255.255 --rate 10000 
\end{Verbatim}

Masscan also forces the following options by default: -sS, -Pn, -n, --randomize-hosts, -v, and --send-eth.

\subsection{Detail Operation of DoH Methods}

\paragraph{DoH-JSON / DoH2-JSON}

This DoH operation mode encodes the parameters of a DNS query as JSON. For transfer it uses HTTPS with the GET method.
The request should enforce the MIME type \texttt{application/dns-json}, because some DoH resolvers do not reply when this value is not present. The reply is a JSON object in plain text. An example usage with \textit{curl} is:

\begin{Verbatim}[fontsize=\footnotesize]
Request:
curl -v -H "Accept: application/dns-json" \
'https://1.1.1.1/dns-query?name=www.example.com&type=A'

Response:
{"Status":0,"TC":false,"RD":true,"RA":true,
"AD":true,"CD":false,
"Question": [{"name":"www.example.com","type":1}],
"Answer":[{"name":"www.example.com","type":1,
"TTL":73014,"data":"93.184.216.34"}]}
\end{Verbatim}

\paragraph{DoH-GET / DoH2-GET}
This DoH operation mode uses HTTPS and the GET method to send the DNS request encoded in BASE64 as a parameter in the URL, a format called \textit{wireformat}. The request should enforce the application/dns-message header because some servers recognize DoH and do not reply when this value is not present. The answer is a sequence of bytes in wireformat.

An example A-type query by \textit{curl} for the domain \url{www.example.com} is:

\begin{Verbatim}[fontsize=\footnotesize]
curl -v -H 'Accept: application/dns-message' 
'https://8.8.8.8/dns-query?dns=q80BAAA...bQAAAQAB'
Response (binary data:
000: abcd8180 00010001 00000000 03777777  .............www
010: 07657861 6d706c65 03636f6d 00000100  .example.com....
020: 01c00c00 01000100 004d1000 045db8d8  .........M...]..
030: 22
\end{Verbatim}

\paragraph{DoH-POST / DoH2-POST}
This mode uses POST method of HTTPS request, and the DNS query is passed as a content of the HTTP request.  and the DNS query is passed as a base64 encoded wireformat message as an HTTP parameter in the URL. The request should enforce accepted \texttt{application/dns-message} because some servers recognize DoH and do not reply when this value is not present. The output is similar to the \textit{DoH-GET / DoH2-GET} example.

An example by \textit{curl} is:

\begin{Verbatim}[fontsize=\footnotesize]
echo "q80BAAABAAAAAAAAA3d3dwdleGFtcGxlA2NvbQAAAQAB" | \ 
base64 -d | curl -v -X POST -H \ 
'Accept: application/dns-message' -H \ 
'Content-Type: application/dns-message' \ 
--data-binary @/dev/stdin 'https://1.1.1.1/dns-query'
\end{Verbatim}

\end{document}